\documentclass[article,nofootinbib,notitlepage]{revtex4-1}

\usepackage[dvips]{graphicx}
\usepackage{amsthm,latexsym,amssymb,amsfonts,array}

\begin{document}

\title{Two-dimensional solutions for Born-Infeld fields}
\author{Rafael Ferraro\medskip}\email{ferraro@iafe.uba.ar}
\thanks{member of Carrera del Investigador Cient\'{\i}fico (CONICET,
Argentina). }

\affiliation{Instituto de Astronom\'\i a y F\'\i sica del Espacio,
Casilla de Correo 67, Sucursal 28, 1428 Buenos Aires, Argentina}
\affiliation{Departamento de F\'\i sica, Facultad de Ciencias
Exactas y Naturales, Universidad de Buenos Aires, Ciudad
Universitaria, Pabell\'on I, 1428 Buenos Aires, Argentina
\bigskip }

\begin{abstract}
The non-linear second order Born-Infeld equation is reduced to a
simpler first order complex equation, which can be trivially solved
for the coordinates as functions of the field. Each solution is
determined by the choice of a holomorphic function subjected to
boundary conditions. The explanation of the method is accompanied by
applications to Born-Infeld electrostatics, magnetostatics and wave
propagation.
\end{abstract}

\maketitle

\section{\label{intro}Introduction}
Born-Infeld equation,
\begin{equation}
(1\,-\,b^{-2}\,c^{-2}\,u_{t}^{2})\ u_{xx}\
+2\,b^{-2}\,c^{-2}\,u_{t}\ u_{x}\ u_{tx}-(1\,+\,b^{-2}\,u_{x}^{2})\
c^{-2}\,u_{tt}=\ 0\, ,  \label{eq0}
\end{equation}%
($u_x$ stands for $\partial_x u$, etc.) is a non-linear wave
equation which is derived from the Lagrangian $L=\sqrt{1-b^{-2}
\left(c^{-2} u_{t}^2-u_{x}^2\right)}$ and appears in several
physical contexts. It descends from the Nambu-Goto action for a
string in 2+1 dimensions when a proper parametrization is chosen
\cite{Bordemann}. Besides, its solutions can be mapped into
solutions of the Galileo-invariant Chaplygin gas in 1+1 dimensions,
since Chaplygin gas is another descendent of the Nambu-Goto action
\cite{Bordemann1, Jackiw}. Born-Infeld equation (\ref{eq0}) also
takes part in Born-Infeld electrodynamics \cite{Born, Born1} when
electromagnetic waves depending just on two variables are
considered. Born-Infeld equation is integrable \cite{Barbashov,
Fairlie} and has a multi-Hamiltonian structure \cite{Arik}; the
corresponding Cauchy problem is studied in
Ref.~\onlinecite{Barbashov1}.

\medskip

Eq.~(\ref{eq0}) is very close to the quasilinear elliptic equation
\begin{equation}
(1\,+\,b^{-2}\,u_{y}^{2})\ u_{xx}\ -2\,b^{-2}\,u_{x}\ u_{y}\
u_{xy}+(1\,+\,b^{-2}\,u_{x}^{2})\ u_{yy}=\ 0\, .  \label{eq00}
\end{equation}
This equation was firstly obtained by Lagrange in 1762 when he
studied the problem of minimizing the area of a surface whose
boundary is a given closed curve in $\mathbb{R}^3$ (Plateau's
problem) \cite{Courant, Xin}; such problem is a natural
generalization of the problem of geodesics. In fact, the {\it
minimal surface equation} (\ref{eq00}) comes from the Lagrangian
$L=\sqrt{1+b^{-2}\left(u_{x}^2+u_{y}^2\right)}$, so the action is
the area of the surface $\zeta = b^{-1}\, u(x,y)$ ($\zeta$ is the
third Cartesian coordinate).\footnote{Vectors $\vec{\delta A}=(dx,\,
0,\, b^{-1}\, u_{x}\, dx)$ and $\vec{\delta B}=(0,\, dy,\, b^{-1}\,
u_{y}\, dy)$ are tangent to the surface $\zeta = u(x, y)$. Thus, the
vector product $\vec{\delta A}\times\vec{\delta B}$ defines the area
of the surface immersed in $\mathbb{R}^3$. The infinitesimal area
then is $|\vec{\delta A}\times\vec{\delta B}|=\sqrt{1+b^{-2}\
\left(u_{x}^2+u_{y}^2\right)}\ dx\, dy$, which leads to the action
for the Eq.~(\ref{eq00}).} Eq.~(\ref{eq00}) says that minimal
surfaces have vanishing mean curvature. The solutions of
Eq.~(\ref{eq00}) can be expressed through a parametric
representation where each solution is determined by the choice of a
pair of related complex functions (Weierstrass-Enneper
parametrization \cite{Dierkes}). The Dirichlet problem for the
Eq.~(\ref{eq00}) is studied in Ref.~\onlinecite{Finn}.

\medskip

Eq.~(\ref{eq00}) could be regarded as a deformed Laplace equation.
Also the equation
\begin{equation}
(1\,-\,b^{-2}\,u_{y}^{2})\ u_{xx}\ +2\,b^{-2}\,u_{x}\ u_{y}\
u_{xy}+(1\,-\,b^{-2}\,u_{x}^{2})\ u_{yy}=\ 0\ ,  \label{eq11}
\end{equation}
which appears in two-dimensional Born-Infeld electrostatics, is a
deformed Laplace equation. Eq.~(\ref{eq11}) is derived from the
Lagrangian $L=\sqrt{1-b^{-2}\left(u_{x}^2+u_{y}^2\right)}$. Their
solutions were characterized by Pryce \cite{Pryce, Pryce1} through a
complex method where each solution is associated with a holomorphic
function (see also References \cite{Ferraro, Ferraro1, Ferraro2}).
Eq.~(\ref{eq11}) is the equation for {\it maximal surfaces}, which
are space-like surfaces in (2+1) Minkowski space with vanishing mean
curvature. Maximal surfaces defined on a domain $D$ of the complex
plane $\varsigma$ also admit a Weierstrass-Enneper parametrization
\cite{Kobayashi}:
\begin{equation}
\left(x(\varsigma),\ y(\varsigma),\ \zeta(\varsigma)\right) = {\it
Re}\ \int \left(\frac{1}{2}f(1+g^2),\ \frac{i}{2}f(1-g^2),\
-fg\right)\ d\varsigma\ ,\label{eq100}
\end{equation}
where $f$ is holomorphic and $g$ is meromorphic on $D$ such that
$fg^2$ is holomorphic on $D$ and $|g(\varsigma)|\neq 1$ for
$\varsigma\in D$.

\medskip

Since the solutions of each one of the equations (\ref{eq0}),
(\ref{eq00}) and (\ref{eq11}) can be transformed into the others by
properly renaming the variables and $b^2$, we will focus just on the
Eq.~(\ref{eq11}). In the following sections we will explain the
method for finding the solutions of Eq.~(\ref{eq11}). We will show
in a few steps that the equation governing the two-dimensional
electrostatic Born-Infeld field can be put into the compact form
(\ref{eq114}), where $e$ is an auxiliary complex field associated
with the real 1-form ${\bf E}\equiv du$, and $z, \overline{z}$ are
complex coordinates ($z=x+i\, y$). In Section \ref{S4} we will
connect the solutions of Eqs.~(\ref{eq0}), (\ref{eq00}) and
(\ref{eq11}) to  field configurations of Born-Infeld
electrodynamics.

\bigskip

\section{\label{S2}Born-Infeld Lagrangian}
The Born-Infeld Lagrangian density for a scalar field is
\begin{equation}
\mathcal{L}[u]\ =\ \sqrt{|g|}\ \sqrt{1-b^{-2}\ g^{kj}\ \partial
_{k}u\
\partial _{j}u}  \label{eq1}
\end{equation}
So the Lagrange equation is
\begin{equation}
\partial _{i}\left( \frac{\sqrt{|g|}\ g^{ij}\ \partial _{j}u}{\sqrt{1-b^{-2}\ g^{kj}\
\partial _{k}u\ \partial _{j}u}}\right) \ =\ 0\ .
\label{eq2}
\end{equation}
We introduce the related 1-forms
\begin{equation}
E_{j}\equiv \partial _{j}u\ ,\hspace{3cm}D_{j}\equiv
\frac{E_{j}}{\sqrt{1-b^{-2}\ g^{kl}\ E_{k}\ E_{l}}}\ .  \label{eq3}
\end{equation}
According to Eq.~(\ref{eq2}) the field $D_{j}$ accomplishes the
equation
\begin{equation}
\partial _{i}\left( \sqrt{|g|}\ g^{ij}\ D_{j}\right) \ =\ 0\ .  \label{eq4}
\end{equation}
In geometric notation, the dynamics of the field is summarized in
the equations \footnote{$\ast $ is the Hodge operator which converts
the 1-form $\mathbf{D}$ into a $(n-1)$-form ($n$ is the dimension).
If $\alpha _{i_{1}.......i_{p}}$ are the components of the $p$-form
$\mathbf{\alpha}$ then $\ast \alpha _{\mu _{p+1}.......\mu
_{n}}=\frac{1}{\,p!}$ $\, \sqrt{|\det (g_{\mu \nu })|}$ $\
\varepsilon _{\mu _{1}.......\mu _{p}\;\mu _{p+1}.......\mu
_{n}}\;\alpha ^{\mu _{1}.......\mu _{p}}$ where $\varepsilon $ is
the Levi-Civita symbol whose value is $1$ ($-1$) for even (odd)
permutations of the natural order of its indexes and vanishes for
repeated indexes.}
\begin{equation}
d\mathbf{E}=0\ ,\hspace{3cm}d\ast \mathbf{D}=0\ , \label{eq5}
\end{equation}
where the 1-forms $\mathbf{E}$ and $\mathbf{D}$ accomplish the
constitutive relation
\begin{equation}
\mathbf{D}\equiv \frac{\mathbf{E}}{\sqrt{%
1-b^{-2}\ \left\Vert \mathbf{E}\right\Vert ^{2}}}\ .\label{eq55}
\end{equation}
One should solve the system (\ref{eq5}), (\ref{eq55})  and then
retrieve the scalar potential $u$ from $\mathbf{E}=du$.

Remarkably, the constitutive relation (\ref{eq55}) is automatically
fulfilled if $\mathbf{E}$ and $\mathbf{D}$ are written in the
suggestive form
\begin{equation}
\mathbf{E}\ =\ \frac{\mathbf{e}}{1+\frac{\left\Vert
\mathbf{e}\right\Vert ^{2}}{4b^{2}}}\ ,\hspace{3cm}\mathbf{D}\ =\
\frac{\mathbf{e}}{1-\frac{\left\Vert \mathbf{e}\right\Vert
^{2}}{4b^{2}}}\ ,  \label{eq6}
\end{equation}
where $\mathbf{e}$ is an auxiliary 1-form field. By replacing
Eq.~(\ref{eq6}) in Eq.~(\ref{eq5}) one gets two equations for
$\mathbf{e}$:
\begin{equation}
\left( 1+\frac{\left\Vert \mathbf{e}\right\Vert ^{2}}{4b^{2}}%
\right) \ d\mathbf{e}\ -\ d\left( \frac{\left\Vert
\mathbf{e}\right\Vert ^{2}}{4b^{2}}\right) \wedge \mathbf{e}\ =\ 0\
,\label{eq61}
\end{equation}
\begin{equation}
\left( 1-\frac{\left\Vert \mathbf{e}\right\Vert ^{2}}{4b^{2}}%
\right) \ d\ast \mathbf{e}\ +\ d\left( \frac{\left\Vert
\mathbf{e}\right\Vert ^{2}}{4b^{2}}\right) \wedge \ast \mathbf{e}\
=\ 0\ .\label{eq62}
\end{equation}

\bigskip

\section{\label{S3}Two-dimensional Euclidean Geometry}
In two-dimensional Euclidean geometry,%
\begin{equation}
ds^{2}=dx^{2}+dy^{2}\ ,\hspace{2cm}g_{ij}=\mathrm{diag}\ (1,1)\ ,
\label{eq10}
\end{equation}
the second order equation (\ref{eq2}) becomes the Eq.~(\ref{eq11}).
With regard to the equivalent system of first order equations
(\ref{eq61}), (\ref{eq62}), we can take advantage of the fact that
both $\mathbf{e}$ and $\ast \mathbf{e}$ are 1-forms if $n=2$. Thus,
we can condensate these equations in a sole complex equation
\begin{equation}
d(\mathbf{e}+i\ \ast \mathbf{e)}\,+\,\frac{\left\Vert
\mathbf{e}\right\Vert ^{2}}{4b^{2}}\ d(\mathbf{e}-i\ \ast \mathbf{
e)}\,-\,d\left( \frac{\left\Vert \mathbf{e}\right\Vert
^{2}}{4b^{2}}\right) \wedge ( \mathbf{e}-i\ \ast \mathbf{e)}\ =\ 0\
, \label{eq110}
\end{equation}
Notice that the complex 1-form $\mathbf{e}-i\, \ast \mathbf{e}$ is
quite elemental in the coordinate basis $(dz,\, d\overline{z})$,
where $z\, = \, x+i\ y$:
\begin{equation}
\mathbf{e}-i\ \ast \mathbf{e\ }=\ (e_{x}\ dx+e_{y}\ dy)-i\ (e_{y}\
dx-e_{x}\ dy)\ =\ (e_{x}-i\ e_{y})\ dz\ .  \label{eq111}
\end{equation}
We will call
\begin{equation}
e\ \equiv \ e_{x}-i\ e_{y}\ .\ \label{eq112}
\end{equation}
Therefore, the field equation (\ref{eq110}) becomes
\begin{equation}
-e\ \ d\left(\frac{|e|^{2}}{4b^{2}}\right)\, \wedge\, dz\ +\
d(\overline{e}\ d\overline{z})\ +\ \frac{|e|^{2}}{4b^{2}}\ \ d(e\
dz)\ =\ 0\ , \label{eq113}
\end{equation}
where we replaced $\left\Vert \mathbf{e}\right\Vert
^{2}=e_{x}^{2}+e_{y}^{2}=|e|^{2}=e\overline{e}$. Eq.~(\ref{eq113})
simplifies to
\begin{equation}
d\overline{e}\,\wedge\, d\overline{z}\ =\ \frac{e^{2}}{4b^{2}}\ \
d\overline{e} \,\wedge\, dz\ .  \label{eq114}
\end{equation}
This complex equation summarizes the dynamics of the field. It can
be tackled from a double perspective. If the auxiliary field $e$ is
regarded as a function of $(z, \overline{z})$ (i.e., as a function
of $(x, y)$), then one gets
\begin{equation}
\frac{\partial \overline{e}}{\partial z}\ =\ -\frac{e^{2}}{4b^{2}}\
\ \frac{\partial \overline{e}}{\partial \overline{z}}\ .
\label{eq20}
\end{equation}
Instead, if the coordinate $z$ is regarded as a function of $(e,
\overline{e})$ one obtains
\begin{equation}
\frac{\partial \overline{z}}{\partial e}\ =\ \frac{e^{2}}{4b^{2}}\ \
\frac{\partial z}{\partial e}\ .  \label{eq21}
\end{equation}
Remarkably, this last form is linear in $z(e, \overline{e})$, and
can be rewritten as
\begin{equation}
\frac{\partial \overline{z}}{\partial \xi }\ =\ -\frac{\partial
z}{\partial (1/\xi )}\ ,\hspace{2cm}\xi\ \equiv\ \frac{e}{2b}\ .
\label{eq24}
\end{equation}
The general solution has the form%
\begin{equation}
z\ =\ f(1/\xi )\ +\ g(\overline{\xi})\ ,  \label{eq25}
\end{equation}
with%
\begin{equation}
g^{\prime }(\xi )\ =\ - f^{\prime }(1/\xi )  \label{eq26}
\end{equation}
(the prime means derivative with respect to the argument).

Eq.~(\ref{eq5}) says that $\mathbf{D}$ is singular when $\left\Vert
\mathbf{E}\right\Vert =b$ (i.e., when $|\xi |=1$). Let us study the
behavior of the solution (\ref{eq25}), (\ref{eq26}) at these values
of the field. We want to know whether $|\xi |=1$ happens at isolated
points or not. For this, we will evaluate $dz$ at $\xi =\exp
[i\ \theta ]$; since%
\begin{equation}
dz\ =\ f^{\prime }(1/\xi )\ d(1/\xi )\ +\ g^{\prime }(\overline{\xi
})\ d(\overline{\xi })\ =\ -g^{\prime }(\xi )\ d(1/\xi )\ +\
g^{\prime }(\overline{\xi })\ d(\overline{\xi })\ ,  \label{eq27}
\end{equation}
then,
\begin{equation}
dz|_{\xi =\exp [i\ \theta ]}=\left[ -g^{\prime }(\exp [i\ \theta
])+g^{\prime }(\exp [-i\ \theta ])\right] \ d(\exp [-i\ \theta ])\ .
\label{eq270}
\end{equation}
Therefore the singularities occurs at isolate points (i.e., $dz=0$) whenever
it is%
\begin{equation}
{\it Im}[g^{\prime }(\xi )]_{|\xi |=1}=0\ .  \label{eq28}
\end{equation}
In Ref.~\onlinecite{Ferraro2}, the reality condition (\ref{eq28})
has been carried out by choosing functions $g^{\prime }$ that do not
change under the transformation $\xi \longrightarrow 1/\xi $ (for
instance, any function with real coefficients that depends just on
$\xi ^{\alpha }+1/\xi ^{\alpha }$). In fact, $\xi $ and $1/\xi $ are
complex conjugate if $\xi =\exp [i\theta ]$; so, such functions are
automatically real on the circle $|\xi |=1$. Even so, it could
happens that ${\it Im}[g^{\prime }(\exp [i\ \theta ])]$ results
ill-defined for some values of $\theta $. In such cases $dz$ in
Eq.~(\ref{eq270}) could be non-null for such particular field
directions (see the multipolar structures in
Ref.~\onlinecite{Ferraro2}). If $g^{\prime }(1/\xi )=g^{\prime }(\xi
)$, then the Eq.~(\ref{eq26}) means that $f^{\prime }(\xi
)=-g^{\prime }(\xi )$; so the solution of Eq.~(\ref{eq24}) becomes
\begin{equation}
z=-g\left( 1/\xi \right) +g\left( \overline{\xi }\right) +constant\ ,
\label{eq280}
\end{equation}
where $g(\xi )$ is any holomorphic function such as
\begin{equation}
g^{\prime }(\xi )=g^{\prime }(1/\xi )\ .  \label{eq281}
\end{equation}
The choice of $g(\xi)$ is constrained by boundary conditions; for
instance, it can be required that $e$ goes to zero for $z$ going to
infinity. Moreover, in Ref.~\onlinecite{Ferraro2} it is shown that
the single-valuedness of the solution $z(\xi ,\overline{\xi})$ and
the isolation of the singular points are guaranteed by functions
$g^{\prime }(\xi )$ which do not possess branch cuts reaching the
circle $|\xi |=1$.

\medskip

\subsection{Complex potential}
The scalar field $u$ is a potential for the field $\mathbf{E}$.
Besides, the closed 1-form $\ast \mathbf{D}$ (see Eq.~(\ref{eq5}))
can be also associated with a potential $v$,
\begin{equation}
\ast \mathbf{D}\ \equiv \ -dv\ .  \label{eq70}
\end{equation}
Then, the complex potential%
\begin{equation}
w\equiv \ u\ +\ i\ v\   \label{eq9}
\end{equation}
satisfies%
\begin{equation}
dw\ =\ \mathbf{E}\ -\ i\ \ast \mathbf{D}\ .  \label{eq8}
\end{equation}
Using Eqs.~(\ref{eq6}) and (\ref{eq111}) one obtains%
\begin{equation}
(2b)^{-1}\ dw\ =\ \frac{\xi \ dz-|\xi |^{2}\ \overline{\xi }\
d\overline{z}}{1-|\xi|^{4}}\ .  \label{eq80}
\end{equation}
Eq.~(\ref{eq114}) implies that $de\wedge dw=0$, which means that $w$
is a holomorphic function of $e$: $w=w(e)$. In fact, using
Eq.~(\ref{eq280}) one gets
\begin{equation}
dw\ =\ \frac{2b}{\xi }\ g^{\prime }(\xi )\ d\xi \ .  \label{eq29}
\end{equation}
Since the function $g^{\prime }(\xi )$ accomplishes the
Eq.~(\ref{eq281}),
the complex potential satisfies%
\begin{equation}
w(\xi )\ =\ -w(1/\xi )\ +\ constant\ .  \label{eq282}
\end{equation}

\medskip

\subsection{\label{S3B}Example: field between grounded conductors}
As explained in the previous subsection, each solution $u(x,y)={\it
Re}[w]$ of Eq.~(\ref{eq11}) is determined by the choice of the
holomorphic function $g^{\prime }(\xi )$ or, alternatively, the
holomorphic function $w(\xi )$; both functions are related by the
Eq.~(\ref{eq29}). We will illustrate the procedure with an example
differing from those considered in Ref.~\onlinecite{Ferraro2}. Let
us choose the function
\begin{equation}
g^{\prime }(\xi )=\frac{d}{\pi }\ \frac{\xi }{1+\xi ^{2}}\ ,  \label{eq310}
\end{equation}
where $d$ is a constant with units of distance. $g^{\prime }(\xi )$
accomplishes the Eq.~(\ref{eq281}). The complex potential in
Eq.~(\ref{eq29}) becomes
\begin{equation}
w\ =\ \frac{2bd}{\pi }\ \arctan [\xi ]\ .  \label{eq311}
\end{equation}
Thus the solution (\ref{eq280}) yields
\begin{equation}
\frac{2\pi }{d}\ z\ =\ \mathrm{Log}\left[ \frac{1+\overline{\xi
}^{2}}{1+1/\xi ^{2}}\right] \ .  \label{eq313}
\end{equation}
This result can be solved for $\xi ^{2}$:
\begin{equation}
\xi^{2}\ =\ -\frac{\exp \left[ \frac{2\pi }{d}\overline{z} \right]
-1}{\exp \left[ -\frac{2\pi }{d}z\right] -1}\ . \label{eq314}
\end{equation}
This field reaches the maximum value $|\xi| = 1$ on the $y$-axis. In
fact if $z = i\ y$ one gets $\xi ^{2} = -1$. The fact that the
maximum value is reached not at an isolated point but on a line is
due to the poles the function $g^{\prime}(\xi)$ displays on the
circle $|\xi| = 1$.

We can also realize that the lines $y = \pm \ d/2$ are equipotential
lines. In fact, by replacing $z = x\pm  i\ d/2$ one gets $\xi^{2} <
0$:
\begin{equation}
\xi ^{2}\ =\ -\frac{\exp \left[ \frac{2\pi }{d}x\right] +1}{\exp
\left[ -\frac{2\pi }{d}x\right] +1}\ ;
\end{equation}
this means that the field is $\xi = -i\xi _{y}$, so it is normal to
the lines $y = \pm \ d/2$. One can check in Eq.~(\ref{eq311}) that
$u = {\it Re}[w]$ vanishes when $\xi $ is pure imaginary.
\begin{figure}[b]
\centerline{\includegraphics[width=3.25in,height=2.00in]{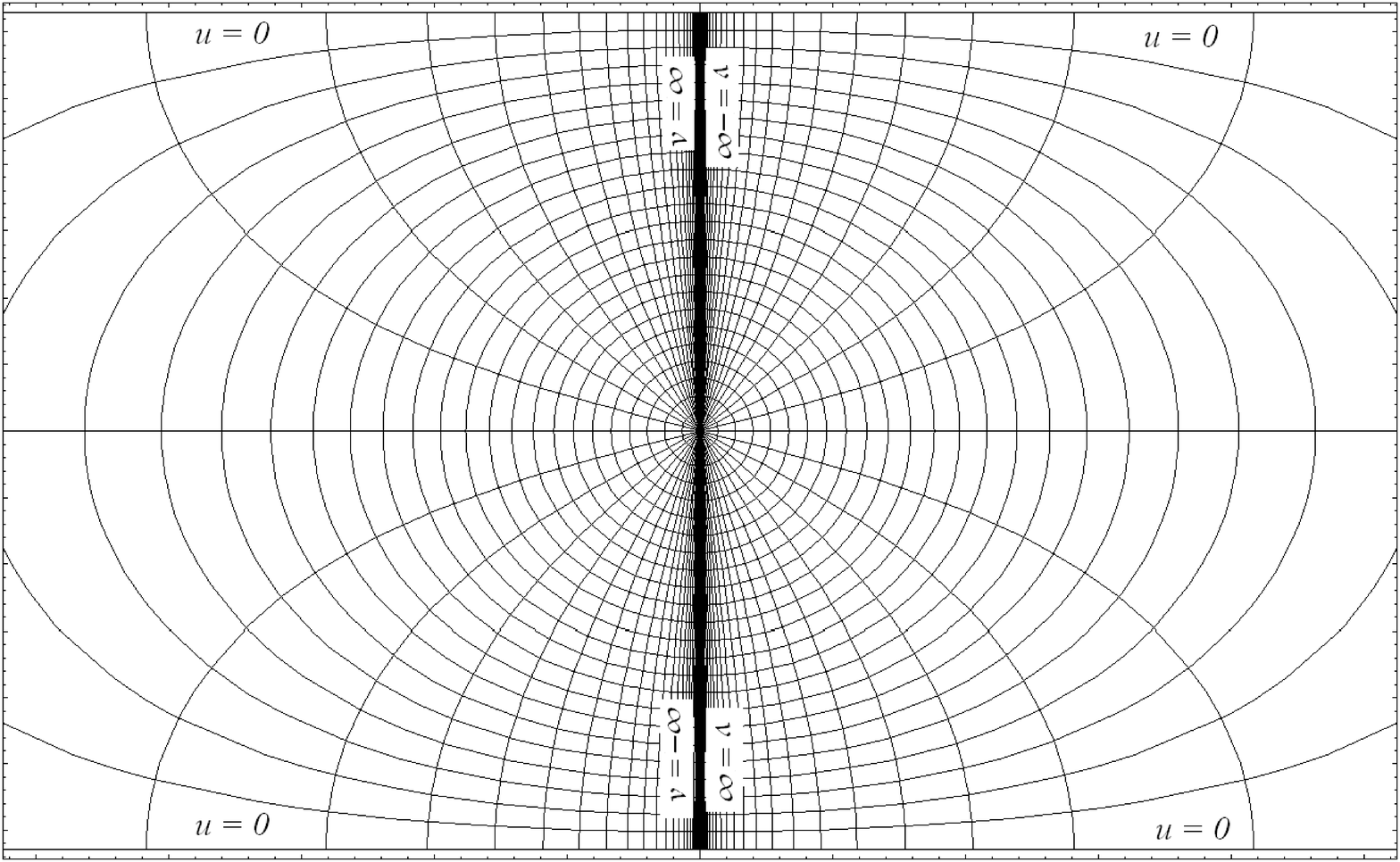}}
\caption{ Equipotential lines and field lines for the field of
Eq.~(\ref{eq314}).} \label{fig1}
\end{figure}
To make a figure of the equipotential lines, we replace $\xi
(w)=\tan [w\pi /(2bd)]$ in Eq.~(\ref{eq313}), with $w=u_{o}+iv$;
then $v$ plays the role of a parameter for the line $u=u_{o}$. Some
equipotential and field lines are shown in Figure 1. Actually, only
the $x\leq 0$ region should be considered in the solution
(\ref{eq314}) since $|\xi |>1$ for $x>0$. However, in Figure 1 the
field has been continuously extended to the semi-space $x>0$, by
choosing $g^{\prime }(\xi )=-(d/\pi )\ \xi /(1+\xi ^{2})$ in this
region. The change of sign in $g^{\prime }(\xi )$ implies the change
$z\longrightarrow -z$ in the solution (\ref{eq314}) and a change of
sign in the expression (\ref{eq311}) for the complex potential.

Figure 1 is the field of a point-like charge between two parallel
grounded conductors separated by a distance $d$. It could also be
regarded as the field of a succession of alternating charges at a
distance $d$ on the $y$-axis. These image charges are joined by
lines of maximum field, as it happens in the multipolar solutions
studied in Ref.~\onlinecite{Ferraro2}. The fact that $v$ ranges
between $-\infty $ and $\infty $ implies that the point-like charge
at the origin is infinite, which is also a characteristic of the
multipolar solutions.\footnote{The charge is the flux of
$\ast\mathbf{D}$: $2\pi Q=\oint D_{x}\,dy-D_{y}\,dx=\oint dv$.}

The field $\mathbf{E}(x,y)=du$ can be computed by differentiating
$u={\it Re}[w(\xi(z,\overline{z}))]$ or directly replacing
$\mathbf{e}$ in Eq.~(\ref{eq6}) with the result (\ref{eq314}) for
the complex auxiliary field.

\medskip

\subsection{\label{S3C}Approximate solution}
In general, we will hardly invert Eq.~(\ref{eq280}) to get an
expression like (\ref{eq314}) for the field $\xi (z,\overline{z})$.
However we could approach the field $\xi (z,\overline{z})$ by
iterating the solution of Eq.~(\ref{eq20}). At the lowest order in
$b^{-2}$ the solution is $e=F(z)$, where $F$ is an analytic
function. It is easy to check that the following order is
\begin{equation}
e(z,\overline{z})=F(z)-\frac{1}{4b^{2}}\ \frac{\partial F(z)}{\partial z}\
\int F(\overline{z})^{2}\ d\overline{z}+\mathcal{O}(b^{-4})\ .  \label{eq301}
\end{equation}
At this order of approximation, the complex potential
$w(z,\overline{z})$ is obtained from Eq.~(\ref{eq80}):
\begin{equation}
dw=e\ dz-\frac{\overline{e}|e|^{2}}{4b^{2}}\
d\overline{z}+\mathcal{O}(b^{-4})\ .  \label{eq303}
\end{equation}
Therefore
\begin{equation}
w(z,\overline{z})=\int F(z)\ dz-\frac{1}{4b^{2}}\ F(z)\ \int
F(\overline{z})^{2}\ d\overline{z}+\mathcal{O}(b^{-4})\ .
\label{eq304}
\end{equation}

\bigskip

\section{\label{S4}Born-Infeld electrodynamics}
Born-Infeld electrodynamics is a non-linear extension of Maxwell
electromagnetism \cite{Born, Born1}. In both theories the
electromagnetic field is an exact 2-form $F=dA$ in Minkowski
spacetime, where the 1-form $A$ is the electromagnetic potential; so
it is $dF=0$. But, differing from Maxwell's field, Born-Infeld
electromagnetic field is governed by the dynamical equations
\begin{equation}
d\ \left(\frac{\ast F+\frac{P}{b^{2}}\
F}{\sqrt{1+\frac{2S}{b^{2}}-\frac{P^{2}}{b^{4}}}}\right)\ =\ 0\ ,
\label{eq32}
\end{equation}
where $S$ and $P$ are the invariants
\begin{equation}
S\ \equiv\ \frac{1}{4}\ F_{ij}\ F^{ij}\ =\ \frac{1}{2}\
(|\overrightarrow{B}|^{2}-|\overrightarrow{E}|^{2}),  \label{eq33}
\end{equation}
\begin{equation}
P\ \equiv\ \frac{1}{4}\ \ast \hfill \!{}F_{ij}\ F^{ij} \ =\
\overrightarrow{E} \cdot \overrightarrow{B}.  \label{eq34}
\end{equation}
The dynamical equations can be derived from the Lagrangian density
\begin{equation}
\mathcal{L}[A]\ =\ \frac{b^{2}}{4\,\pi }\;\sqrt{|g|}\
\left(1-\sqrt{1+\frac{2S}{b^{2}}-\frac{P^{2}}{b^{4}}}\right) \ .
\label{eq35}
\end{equation}
Notice that Maxwell's theory is recovered in the limit
$b\longrightarrow \infty $. Born-Infeld Lagrangian (\ref{eq35}) is
exceptional because, together with another unphysical Lagrangian, is
the only function of $S$ and $P$ ensuring the absence of
birefringence and shock waves \cite{Boillat, Plebanski, Bialynicki,
Kerner}.

Except for the field of a point-like charge \cite{Born1} and the
essentially two-dimensional solutions we are going to show in this
Section, it is very hard to find exact solutions for Born-Infeld
electrodynamics. Maxwell's plane waves are trivial solutions because
they have vanishing invariants $S$ and $P$; so no difference remains
between Maxwell and Born-Infeld equations in such case. The exact
solution for a plane wave interacting with a static uniform field
has been obtained in Ref.~\onlinecite{Aiello}. The case for a
cylindrical wave has been recently worked out \cite{Petrov}.
Stationary solutions were studied under the form of perturbative
series \cite{Kiessling}; the uniqueness of such solutions was also
examined \cite{Kiessling1}. It has been shown that Born-Infeld
dynamics can be thrown into a form similar to MHD equations by
promoting the Poynting vector and the energy to the status of
unknown variables \cite{Brenier}. The chance of detecting effects of
Born-Infeld electrodynamics in laser-plasma experiments is analyzed
in Refs.~\onlinecite{Dereli, Burton}.

As we will show in this Section, the Eqs.~(\ref{eq0}), (\ref{eq00})
and (\ref{eq11}) describe some Born-Infeld field configurations with
$P=0$. By replacing $P=0$ in Eq.~(\ref{eq32}), the dynamical
equations become
\begin{equation}
\left( 1+b^{-2}\ 2S\right) \ d\ast F-b^{-2}\ dS\wedge \ast F\ =\ 0\ .
\label{eq36}
\end{equation}

\medskip

\subsection{\label{S4A}Pure electric field}
Let us consider the electromagnetic potential
\begin{equation}
A\ =\ u(x,y)\ c\,dt\ .  \label{eq37}
\end{equation}%
Then $F = u_{x}\ c\,dt\wedge dx + u_{y}\ c\,dt\wedge dy$ is an
electrostatic field whose field lines lie on the $(x,y)$-plane. The
electric components of the field are $F_{0\alpha} = E_{\alpha} =
\partial_{\alpha}u$. Thus,
\begin{equation}
2S=-u_{x}^{\;2}-u_{y}^{\;2}\ ,  \label{eq38}
\end{equation}
\begin{equation}
\ast F=u_{x}\ dy\wedge d\zeta-u_{y}\ dx\wedge d\zeta\ \label{eq39}
\end{equation}
($\zeta$ is the third Cartesian coordinate). In this case the
dynamical equation (\ref{eq36}) turns out to be the
Eq.~(\ref{eq11}). This is because the Lagrangian (\ref{eq35}) is
essentially $\sqrt{1-b^{-2}|\overrightarrow{E}|^{2}}$, so it
coincides with the Lagrangian in Section \ref{S3}.

The simplest example is the cylindrically symmetric field associated
with a charge density $\lambda$ distributed along the $\zeta$-axis.
According to the definition of $e$ (see Eq.~(\ref{eq112})), such a
radial symmetry requires that $\arg (z) = -\arg (\xi )$. Then,
$g(\xi)$ in Eq.~(\ref{eq280}) has to be linear; thus $g^{\prime
}(\xi )$ is a real constant and accomplishes the reality condition
(\ref{eq28}). The value of the constant $g^{\prime}$ is dictated by
the Coulombian limit $b\rightarrow\infty$; we will see that
$g^{\prime} = \lambda /(2b)$. Therefore the cylindrically symmetric
solution (\ref{eq280}) is
\begin{equation}
z\ =\ \frac{\lambda }{2b}\left( \frac{1}{\xi }-\overline{\xi
}\right) \ . \label{eq41}
\end{equation}
In this case, the function $z(\xi,\overline{\xi})$ is easily
inverted to obtain
\begin{equation}
\xi (z,\overline{z})\ =\ \frac{b}{\lambda }\ \overline{z}\ \left(
\sqrt{1+\frac{\lambda ^{2}}{b^{2}\,z\,\overline{z}}}-1\right)\ =\
\frac{\lambda }{b\,z}\ \left( \sqrt{1+\frac{\lambda
^{2}}{b^{2}\,z\,\overline{z}}}+1\right) ^{-1}\ . \label{eq42}
\end{equation}
To compute the potential $u(x,y)$, let us integrate the
Eq.~(\ref{eq29}) for $g^{\prime} = \lambda /(2b)$; it results
\begin{equation}
w(\xi)\ =\ -\lambda\ \mathrm{Log}[\xi]\, .\label{eq40}
\end{equation}
Then, by replacing the expression (\ref{eq42}) one gets
\begin{eqnarray}
u(x,y)\ &=&\ {\it Re}\left\{ \lambda \,\mathrm{Log}\left[
\frac{\lambda }{b\ (x+i\,y)}\ \right] -\lambda \,\mathrm{Log}\left[
\sqrt{1+\frac{\lambda ^{2}}{b^{2}(x^{2}+y^{2})}}+1\right] \right\}
\nonumber \\  \nonumber \\
&=&\ -\lambda \ \mathrm{\log }\left[ \sqrt{b^{2}\lambda
^{-2}r^{2}+1}+|\lambda |^{-1}b\ r\right] \ ,  \label{eq43}
\end{eqnarray}%
where $r^{2}=x^{2}+y^{2}$. When $b\longrightarrow \infty $ one
recovers the Coulombian potential $u=-\lambda \ \mathrm{\log
}[r]+$\textit{constant}.

\medskip

\subsection{Pure magnetic field}
The electromagnetic potential
\begin{equation}
A=u(x,y)\ d\zeta  \label{eq44}
\end{equation}
($\zeta $ is the third Cartesian coordinate) leads to a pure
magnetic field whose field lines lie on the $(x,y)$-plane. In fact,
the field $F$ is deprived of components $F_{0\alpha }$: $F = u_{x}\
dx\wedge d\zeta + u_{y}\ dy\wedge d\zeta $. Since $B_{\alpha} =
\epsilon_{\alpha \beta\gamma }F^{\beta \gamma }$, then it is $B_{x}
= u_{y}$ and $B_{y} = -u_{x}$. Thus,
\begin{equation}
2S\ =\ u_{x}^{\;2}\ +\ u_{y}^{\;2}\ ,  \label{eq45}
\end{equation}
\begin{equation}
\ast F\ =\ -u_{x}\ c\,dt\wedge dy\ +\ u_{y}\ c\,dt\wedge dx\ .
\label{eq46}
\end{equation}
The dynamical equation (\ref{eq36}) becomes Eq.~(\ref{eq00}), which
differs from Eq.~(\ref{eq11}) in the sign of $b^{-2}$; this is
because now the Lagrangian is basically
$\sqrt{1+b^{-2}|\overrightarrow{B}|^{2}}$ instead of
$\sqrt{1-b^{-2}|\overrightarrow{E}|^{2}}$. Any solution $u(x,y)$
obtained through the procedure explained in the previous Section can
be converted in a solution of Eq.~(\ref{eq00}) by changing $
b^{2}\longrightarrow -\,b^{2}$. Alternatively, one can also change $
x\longrightarrow i\,x$, $y\longrightarrow i\,y$, which is equivalent
to change $z\longrightarrow i\,z$ and $\overline{z}\longrightarrow
i\,\overline{z}$ in $u(z,\overline{z})={\mathit
Re}[w(z,\overline{z})]$.

For instance, according to Eq.~(\ref{eq43}) the potential for a neutral
straight steady current has the form
\begin{equation}
u\ =\ -\lambda \ \mathrm{\log }\left[ \sqrt{b^{2}\lambda
^{-2}r^{2}-1}+\sqrt{b^{2}\lambda ^{-2}r^{2}}\right] \ =\ -\lambda \
\left\vert\mathrm{arccosh}\left[\frac{b\
r}{\lambda}\right]\right\vert, \label{eq470}
\end{equation}
which represents a catenoid in $\mathbb{R}^3$ (a well known minimal
surface \cite{Xin}). The field $F=(du/dr)\ dr\wedge d\zeta $
diverges at $r=b^{-1}\lambda $. This result could mean that pure
Born-Infeld magnetostatic fields are just an approximation to be
used far from the sources. Near to the sources one should not ignore
the true nature of the charges that constitute the steady current.

Notice that the replacement $b^{2}\longrightarrow -\,b^{2}$ in the
Lagrangian (\ref{eq1}) amounts the exchange of roles between
$\mathbf{E}$ and $\mathbf{D}$ in Eq.~(\ref{eq6}). Thus, the
association between the Born-Infeld magnetostatic field
$B_{x}=u_{y}$, $B_{y}=-u_{x}$ and the complex field $e$ turns out to
be
\begin{equation}
\ -B_{y}-i\ B_{x}\ =\ u_{x}-i\ u_{y}=\
\frac{2b}{\frac{2b}{e}-\frac{\overline{e}}{2b}}\ .  \label{eq471}
\end{equation}
The auxiliary field $e(z,\overline{z})$ is now governed by the
equation
\begin{equation}
\frac{\partial \overline{z}}{\partial e}+\frac{e^{2}}{4b^{2}}\
\frac{\partial z}{\partial e}\ =\ 0\ ,
\end{equation}
whose general solution is
\begin{equation}
z\ =\ f(1/\xi )\ -\ g(\overline{\xi })\
,\hspace{0.25in}\mathrm{where}\hspace{0.25in} -g^{\prime }(\xi )\ =\
f^{\prime }(1/\xi )\ .
\end{equation}
According to Eq.~(\ref{eq471}), $\overrightarrow{B}$ is singular where $|\xi
|=1$.

\medskip

\subsection{\label{S4C}Stationary waves}
Let us start with the electromagnetic potential%
\begin{equation}
A\ =\ u(x,t)\ dy\ .  \label{eq48}
\end{equation}
The field has now electric and magnetic orthogonal components, $F =
c^{-1}\ u_{t}\ c\,dt\wedge dy + u_{x}\ dx\wedge dy$. Then, $E_{y} =
c^{-1}\ u_{t}$ and $B_{\zeta } = u_{x}$. Therefore,
\begin{equation}
2S\ =\ u_{x}^{\;2}-c^{-2}\ u_{t}^{\;2}\ ,  \label{eq49}
\end{equation}
\begin{equation}
\ast F\ =\ c^{-1}\ u_{t}\ dx\wedge d\zeta\ +\ u_{x}\ c\,dt\wedge
d\zeta \ . \label{eq50}
\end{equation}
In this case the dynamical equation (\ref{eq36}) yields the
Born-Infeld equation (\ref{eq0}). This equation becomes
Eq.~(\ref{eq11}) by replacing $y\longrightarrow i\,ct $ and
$b\longrightarrow i\,b$.

We will look for stationary waves between two parallel conductors.
We will apply the expression (\ref{eq301}) to get approximate
solutions. Stationary waves can be obtained by starting from the
holomorphic function $F(z)=e_{o}\cos kz$. In fact, according to
Eq.~(\ref{eq304}) this choice implies the potential
\begin{equation}
u\ =\ {\mathit Re}[w]\ =\ {\mathit Re}\left[-\frac{e_{o}}{k}\ \sin
kz\right]\vert_{y=i\,ct}+\mathcal{O}(b^{-2})=-\frac{e_{o}}{k}\ \cos
kct\ \sin kx+\mathcal{O}(b^{-2})\, .
\end{equation}
So, for $b\longrightarrow \infty $ one obtains the Maxwellian
potential for stationary waves. Let us use the Eq.~(\ref{eq301}) to
compute the next order of approximation:
\begin{equation}
e(z,\overline{z})\ =\ e_{o}\cos kz\ +\ \frac{e_{o}^{3}}{16\, b^{2}}\
\sin kz\ \left( 2k\overline{z} + \sin 2k\overline{z}\right)\ +\
\mathcal{O}(b^{-4})\ . \label{eq52}
\end{equation}
Although we expected an oscillating solution, the approach
(\ref{eq301}) produced a secular term $2k\overline{z}$. This means
that the result (\ref{eq52}) is valid just for
$k|z|e_{o}^{2}b^{-2}<<1$. The secular term can be healed by
replacing $k\overline{z}e_{o}^{2}b^{-2}/8$ with $\sin (k
\overline{z}e_{o}^{2}b^{-2}/8)$. In fact, it is easy to verify that
the field
\begin{eqnarray}
e(z,\overline{z}) &=& e_{o}\cos kz\ +\ e_{o}\sin kz\ \sin
\frac{e_{o}^{2}k \overline{z}}{8\, b^{2}}\ +\ \frac{e_{o}^{3}}{16\,
b^{2}}\ \sin kz\ \sin 2k\overline{z}\ +\ \mathcal{O}(b^{-4})  \nonumber \\
&=& e_{o}\cos k\left( z-\frac{e_{o}^{2}}{8\, b^{2}}\
\overline{z}\right)\ +\ \frac{e_{o}^{3}}{16\, b^{2}}\ \sin kz\ \sin
2k\overline{z}\ +\ \mathcal{O} (b^{-4})\ .\   \label{eq533}
\end{eqnarray}%
accomplishes the Eq.~(\ref{eq20}) at the considered order of
approximation. The secular term then expresses a correction to the
frequency of the stationary wave.

We can use the Eq.~(\ref{eq304}) to compute the complex potential
$w(z, \overline{z})$. After healing the secular term, we get the
complex potential fulfilling the Eq.~(\ref{eq303}) for the field
(\ref{eq533}):
\begin{eqnarray}
w(z,\overline{z}) &=& \frac{e_{o}}{k}\ \sin kz-\frac{e_{o}}{k}\ \cos
kz\ \sin \frac{e_{o}^{2}k\overline{z}}{\ 8b^{2}}-\frac{e_{o}^{3}}{\
16\ k\ b^{2}}\cos
kz\ \sin 2k\overline{z}+\mathcal{O}(b^{-4})\   \nonumber \\
&=& \frac{e_{o}}{k}\ \sin k\left( z-\frac{e_{o}^{2}}{8\, b^{2}}\
\overline{z}\right) -\frac{e_{o}^{3}}{16\ k\ b^{2}}\ \cos kz\ \sin
2k\overline{z}+\mathcal{O}(b^{-4})\ .  \label{eq534}
\end{eqnarray}
We now get the real potential $u(x,y) = {\mathit Re}[w]$ that
accomplishes the Eq.~(\ref{eq11}), and pass to the solution of
Born-Infeld equation (\ref{eq0}) by changing $y\longrightarrow
i\,ct$ and $b\longrightarrow i\,b$:
\begin{eqnarray}
u(x,t) &=& \frac{e_{o}}{2k}\ \sin k\left( x-c\,t+\frac{e_{o}^{2}}{8\
b^{2}}\ (x+c\,t)\right) + \frac{e_{o}}{2k}\ \sin k\left(
x+c\,t+\frac{e_{o}^{2}}{8\
b^{2}}\ (x-c\,t)\right)  \nonumber \\
&& +\frac{e_{o}^{3}}{32\ k\ b^{2}}\ \cos k(x-c\,t)\ \sin 2k(x+c\,t)
+ \frac{e_{o}^{3}}{32\ k\ b^{2}}\ \cos k(x+c\,t)\ \sin
2k(x-c\,t)+\mathcal{O}(b^{-4})\ .
\end{eqnarray}
At the considered order of approximation, the result can be
reorganized as
\begin{eqnarray}
u(x,t) &=& \frac{e_{o}}{k}\ \sin \left[ \left(
1+\frac{e_{o}^{2}}{8\, b^{2}}\right) kx\right] \ \cos \left[ \left(
1-\frac{e_{o}^{2}}{8\, b^{2}}\right) kct\right]  +
\frac{e_{o}^{3}}{16\ k\ b^{2}}\ \sin kx\ \cos kct\ (\cos 2kx+\cos
2kct)+
\mathcal{O}(b^{-4})\ \nonumber \\ \nonumber \\
&=& \frac{e_{o}}{k}\ \sin \left[ \left( 1+\frac{e_{o}^{2}}{8\,
b^{2}}\right) kx \right] \ \cos \left[ \left( 1-\frac{e_{o}^{2}}{8\,
b^{2}}\right) kct\right] \ \left( 1+\frac{e_{o}^{2}}{16\ \ b^{2}}\
(\cos 2kx+\cos 2kct)\right) + \mathcal{O}(b^{-4})\ .
\end{eqnarray}
To fulfill boundary conditions $u(0,t)=0$ and $u(d,t)=0$,
corresponding to two parallel grounded conductors at a distance $d$,
we choose
\begin{equation}
k=\frac{n\pi }{d}\left( 1-\frac{e_{o}^{2}}{8\, b^{2}}\right)
+\mathcal{O}(b^{-4})\ .
\end{equation}
Then, as a consequence of the non-linearity, the resonant
frequencies in a cavity depend on the amplitude.

The obtained solution can be boosted along the parallel conductors
to get propagating waves in a waveguide \cite{Ferraro3, Ferraro4}.

\bigskip

\section{Conclusion}
We have shown a method to build solutions of Born-Infeld equation
(\ref{eq0}) and its relatives (\ref{eq00}), (\ref{eq11}). The method
exploits the power of exterior calculus in the complex basis of
$\mathbb{R}^2$, which is the natural language for this problem.
Although the method was developed for the Eq.~(\ref{eq11}), the
obtained solutions are converted into solutions to the other
equations by properly changing the variables or the Born-Infeld
constant $b^2$.

\medskip

Remarkably, Eq.~(\ref{eq11}) becomes Laplace equation at the points
where the first derivatives $u_{x}$, $u_{y}$ vanish. This
distinctive feature prevents the existence of smooth extremes in
static Born-Infeld configurations. In fact, Laplace equation could
not be fulfilled at an extreme since $u_{xx}$ and $u_{yy}$ should
have the same sign. This means that Born-Infeld dynamics does not
harbor smooth static solutions going to zero at the boundaries
(i.e., no solitary waves exist other than those traveling at the
speed of light). These aspects of extremes in Born-Infeld
electrostatics can be recognized in the examples shown in Sections
\ref{S3B} and \ref{S4A}.

\medskip

Since Eq.~(\ref{eq114}) is linear in $z$, the general solution
(\ref{eq280}) expresses $z$ as a function of the auxiliary complex
field. In most of the cases, it will be very hard to invert this
function for obtaining the field as a function of the coordinates.
However, since non-linear effects are expected to be very weak, just
an expression at the lowest order in $b^{-2}$ would be enough for
experimental tests. This is the case of the approximate expressions
(\ref{eq301}) and (\ref{eq304}) that we applied in Section \ref{S4C}
to study Born-Infeld stationary waves in a cavity.

\end{document}